# A Multi-Step Reconfiguration Model for Active Distribution Network Restoration Integrating DG Start-Up Sequences

Hossein Sekhavatmanesh, *Student Member, IEEE*, Rachid Cherkaoui, *Senior Member, IEEE*

*Abstract*—The ever-increasing penetration of Distributed Generators (DGs) in distribution networks suggests to enable their potentials in better fulfilling the restoration objective. The objective of the restoration problem is to resupply the maximum energy of loads considering their priorities using minimum switching operations. Basically, it is desired to provide a unique configuration that is valid regarding the load and generation profiles along the entire *restorative period*. However, this unique configuration may not satisfy at the same time: I) the DG start-up requirements at the beginning of the restoration plan and II) the topological conditions that would allow the DG to provide later on the most efficient support for the supply of loads. Therefore, it is proposed in this paper to allow a limited number of reconfiguration steps according to the DG start-up requirements. In addition, this paper presents a novel formulation for the reconfiguration problem that accounts for partial restoration scenarios where the whole unsupplied area cannot be restored. The decision variables of the proposed *multi-step* restoration problem are: I) the line switching actions at each step of the reconfiguration process, II) the load switching actions during the whole *restorative period* and, III) the active/reactive power dispatch of DGs during the whole *restorative period*. A relaxed AC power flow formulation is integrated to the optimization problem in order to ensure the feasibility of the solution concerning the operational safety constraints. The overall model is formulated in terms of a mixed-integer second-order cone programming. Two simulation scenarios are studied in order to illustrate different features of the proposed strategy and to demonstrate its effectiveness particularly in the case of large-scale outages in distribution networks.

## NOMENCLATURE

### A. Parameters

| | |
|---|---|
| $w_{re}, w_{sw}, w_{op}$ | Weighting factors of the objective function terms (p.u.) |
| $D_i$ | Importance factor of the load at bus $i$ (p.u.) |
| $\lambda_{ij}$ | The operation time of line switch $ij$ (hour). |
| $P_{i,t}^D (Q_{i,t}^D)$ | Active (Reactive) power demand at bus $i$, at time $t$ (p.u.) |
| $r_{ij} (x_{ij})$ | Resistance (Reactance) of line $ij$ (p.u.). |
| $v^{max} (v^{min})$ | Maximum (Minimum) limits of voltage magnitude (p.u) |
| $f_{ij}^{max}$ | Maximum current flow rating of line $ij$ (p.u) |
| $P_{i,max}^{inj}$ | Active power capacity of resource at node $i$ (p.u.) |
| $Q_{i,max}^{inj} (Q_{i,min}^{inj})$ | Maximum (Minimum) reactive power allowed to inject/extract from resource at node $i$ (p.u.). |
| $S_{i,max}^{inj}$ | Apparent power capacity of resource at node $i$ (p.u.) |
| $A_{ij,k}$ | Indicator specifying if line $ij$ is totally in zone $k$ (1/0) |
| $M$ | A large multiplier |
| $\Delta_i$ | The energization time constant of DG at node $i$ (hour). |
| $\Delta t$ | Time step length (hour) |

### B. Variables

| | |
|---|---|
| $Y_{ij,s}$ | Binary decision variable indicating if at step $s$ the line $ij$ equipped with a switch is energized or not (1/0) |
| $L_{i,t}$ | Binary decision variable indicating if at time $t$ the load at node $i$ is supplied or rejected (1/0) |
| $Z_{ij,s}$ | Continuous variable indicating if at step $s$ the line $ij$ is oriented from node $i$ to node $j$ or not. |
| $\Psi_{ij,s}$ | Auxiliary flow that is travelling at step $s$ in line $ij$ from node $i$ to node $j$. |
| $X_{i,s}$ | Indicator of node $i$ at step $s$, being energized or not (1/0). |
| $X_{ij,s}$ | Indicator of line $ij$ at step $s$, being energized or not (1/0) |
| $S_{ij,s}$ | Continuous variable indicating if at step $s$ sectionalizing switch on line $ij$ will be operated or not (1/0). |
| $T_s$ | The starting time instant of reconfiguration step $s$ (hour). |
| $K_{t,s}$ | Continuous variable indicating if at time $t$ the network is under the configuration at step $s$ or not (1/0). |
| $F_{ij,t}$ | Square of current flow magnitude in line $ij$, at time $t$ (p.u) |
| $P_{ij,t} (q_{ij,t})$ | Active (Reactive) power flow in line $ij$, starting from node $i$, at time $t$ (p.u). |
| $P_{i,t}^{inj} (Q_{i,t}^{inj})$ | Active (Reactive) power injection from the substation or DGs at node $i$, at time $t$ (p.u.). |
| $V_{i,t}$ | Square of voltage magnitude at bus $i$, at time $t$ (p.u). |

### C. Indices

| | |
|---|---|
| $i, j$ | Index of nodes |
| $ij$ | Index of branches |
| $t$ | Index of time |
| $s$ | Index of reconfiguration step |

### D. Sets

| | |
|---|---|
| $N$ | Set of nodes |
| $W$ | Set of lines (plus tie-lines) |
| $N^*$ | Set of nodes in the off-outage area |
| $Z^*$ | Set of zones in the off-outage area |
| $W^*$ | Set of lines (plus tie-lines) in the off-outage area |
| $W^S$ | Set of lines in the off-outage area equipped with tie ($W_{tie}^S$) and sectionalizing switches ($W_{sec}^S$) |
| $W_{ava}^S$ | Set of available tie-lines |
| $\Omega_{Res}$ | Set of injection nodes including substations ($\Omega_{Sub}$) and DGs ($\Omega_{DG}$) |

## I. INTRODUCTION

Nowadays, moving toward the smart grid concept has changed the face of distribution systems. This movement is motivated mostly by the highly increasing penetration of Distributed Generators (DGs) at the distribution system level. One of the main features of smart grids is their self-healing capability thanks to the new generation of monitoring, communication and control facilities. In this regard, in case of fault, the restoration of distribution systems, considering their new active status, is a timely topic that deserves to be revisited.

When a fault occurs in a radial distribution network, once it is isolated, the area downstream to the fault place (*off-outage area*) remains unsupplied even in presence of DGs in that area.

Actually, according to the IEEE standard 1547, following a severe disturbance, every DG in the network should be automatically disconnected [1]. The restoration problem aims to supply the maximum loads in this off-outage area using a minimum number of switching operations. In this respect, the off-outage area will be restored through healthy feeders that can be directly connected to it through normally-open (tie) switches. These feeders and switches are called *available feeders* (Feeder-b in Fig. 1) and *available tie-switches* (T3 in Fig. 1), respectively. In order to maintain the radiality of the network configuration, some of the normally-closed (sectionalizing) switches should be open. The resulting new configuration of the network remains for a so-called *restorative period* starting from the fault isolation instant until the time when the faulted element is repaired. After the restorative period, the original configuration of the network will be restored.

In ADNs, besides the switching operations, the set point modifications of dispatchable DGs can also be integrated into the restoration strategies. In this paper, when the term of DG is used, dispatchable DG is meant. The DG-aided restoration strategies have attracted a lot of attention among researchers in recent years. A group of these studies propose to set up intentional DG-island systems in order to supply locally a group of nodes by at least one DG working in grid-forming mode [2], [3]. Heuristic approaches guided by expert-based rules are used to find the suitable DG-island system regarding the targeted restoration objectives. While these studies account for different DG operational requirements, system requirements are mostly disregarded. These requirements concern among others the protection settings, voltage regulation equipment and settings, and proper operation of loads in the DG-island system. Therefore, for the restoration strategy proposed in this paper, all the DGs are assumed to be operated in the grid-feeding mode only.

According to the common practices in electrical utilities, it is desirable to provide a unique configuration that is valid regarding the load and generation profiles along the entire restorative period. This so-called *static approach* may not be able to handle some dynamic processes that exist during the service restoration. Among these dynamic processes is the DG start-up process, which is neglected in the DG-aided restoration strategies proposed in the literature [4], [5]. This practical concern is the main focus of this paper, It will be defined and discussed in details in section II.

In order to account for a time-dependent process in the reconfiguration problem, the papers in the literature propose another approach, which is called *dynamic approach*. In the dynamic reconfiguration approaches, the topology of the network continually changes specifically in response to the variation of loads and intermittent resources. For instance, the authors in [6] and [7] propose intraday dynamic reconfiguration models with the aim of minimizing daily running costs, and minimizing DG power curtailments, respectively. However, the network configuration should not change frequently during the restorative period. The reason is to avoid many interruptions to the customers by transient switching disturbances especially in the restored part of the network that can be already pushed to the edge of its capabilities. Finally, we can conclude that the dynamic reconfiguration approach is also inappropriate for the restoration problem studied in this paper.

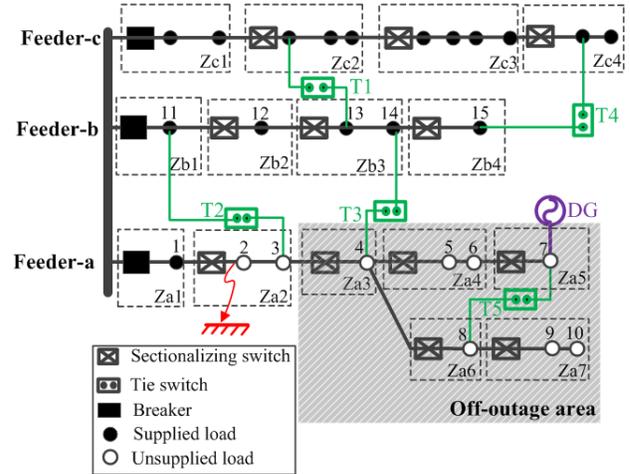

Fig. 1. A simple distribution network under fault conditions

From mathematical programming point of view, the reconfiguration problem is a mixed-integer and non-linear multi-period optimization problem, respectively due to the switching decisions and power flow equations. Therefore, the resulting model leads to an NP-hard combinatorial optimization problem. In order to handle the non-polynomial hardness of such an Optimal Power Flow (OPF)-based optimization problem, most of the papers apply heuristic and meta-heuristic methods such as an improved multi objective harmony search algorithm [8], Multi Objective Molecular Differential Evolution (MOMDE) method [9], and Shuffled Frog Leaping Algorithm (SFLA) [10]. However, these methods are in general time-consuming and do not guarantee to provide high-quality solutions or even to find an existing feasible solution.

The computational complexity of the reconfiguration problem is even higher in the case of dynamic approaches, since they need to integrate many binary variables indicating the switching operations at every time step. In order to relax the computation burden of the dynamic reconfiguration problems, papers [11] and [12] present methodologies to partition the time window of the optimization problem in clusters with similar load levels. The configuration for each cluster remains unchanged. Therefore, the number of required binary variables is reduced and the computation burden is significantly relaxed. However, this clustering approach is not applicable if we plan to account for the DG start-up process. As it will be explained in section II, regarding the restoration problem studied in this paper, the time window (i.e. restorative period) should decompose in clusters at the time instants when the disconnected DGs are energized. These time instants are not pre-determined parameters like those of the variability of load and generations which are the only time-dependent processes studied in [11] and [12].

The authors in [13] propose to solve sequentially the reconfiguration problem for each time step (hour) of the day (successive single period optimization problems). Prior to each time step, an updated forecasting time horizon is considered which is obtained using stochastic model predictive control. In this respect, the dynamic reconfiguration problem is transformed into a series of static reconfiguration problems at different time steps. The weakness of this strategy is that the solutions at a given time step are not influenced by the future decisions during the rest of the optimization process. Therefore, this approach cannot be applied to solve multi-period restoration problems (such as the one studied in this paper), where the feasible and optimal solution at one time step depends on the solution of the problem in the previous and next steps.

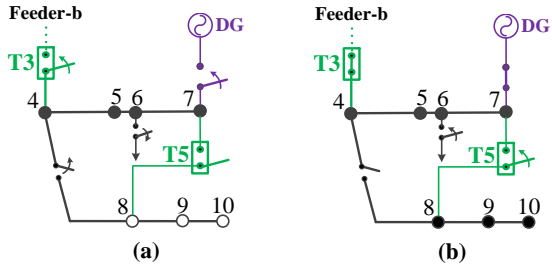

Fig. 2. Two reconfiguration steps of the restoration strategy in case of the fault in the test system of Fig. 1. a) step1, b) step 2.[1]

In this paper, an analytical *multi-step* reconfiguration problem is proposed and applied to the restoration problem. A relaxed AC-power flow formulation is integrated to the optimization problem in order to ensure the feasibility of the solution concerning the operational safety constraints. The overall model is formulated in terms of a Mixed-Integer Second-Order Cone Programming (MISOCP). This formulation addresses the aforementioned weaknesses of static and dynamic reconfiguration approaches handling time-dependent processes existing in the restoration problem. More specifically, we aim in this paper to address the requirements of the DG start-up process in the restoration solution. These requirements together with the proposed multi-step reconfiguration problem are defined in section II and then formulated in section III. Section IV provides two simulation scenarios illustrating the main contributions of this paper. Finally, section V concludes the paper with final remarks concerning the applicability of the proposed multi-step reconfiguration problem.

## II. PROBLEM STATEMENT

In this paper, it is aimed to provide a restoration formulation complying with the DG start-up requirements. This process is defined for DGs that are in the off-outage area. Once the fault occurs, as already said in section I, these DGs should be automatically disconnected [1]. In order to enable these DGs to re-inject power into the network, they should go under a specific process called DG start-up process. The requirements of this DG start-up process that are impacting the restoration solution are defined in two-folds: I) energization of the DG hosting node, and II) start-up duration of the DG. Since in this paper the DGs are assumed to operate only in grid-connected mode, at the beginning of the restoration, the disconnected DGs need a configuration that provides them feasible path(s) from the healthy part of the grid. Once the hosting node of the DG is energized, a sequence of actions is carried out to enable the DG to inject power. These actions take a period called *DG start-up duration*. They involves among others the repose time of the DG primary source and the time for launching auxiliary control systems [14]. The detailed procedure of the DG start-up is not dealt with in this paper. Only the two requirements mentioned above (I and II) are considered and formulated in section III as (28) and (29), respectively.

While the DG is still not ready to inject power, some parts of the off-outage area might need to be left unrestored. Once the DG is ready to produce, a new configuration might be needed in order to supply, if possible, those unrestored areas/loads thanks to a more efficient support of the DG. As an example, consider the faulted network shown in Fig. 1. In the first reconfiguration step shown in Fig. 2.a, the DG hosting node is energized by closing tie-switch T3. Since the DG does not inject power in this step, loads 6, 8, 9, and 10 cannot be restored while respecting the network safety constraints. In Fig. 2.b, the DG is ready to inject power in the grid-connected mode and supports feeder-b in supplying the loads. Therefore, tie-switch T5 and load breaker 6 are closed to supply the loads that were left unrestored in the first reconfiguration step.

Therefore, throughout the restoration process, the off-outage area changes its configurations through a certain sequence of switching called *network reconfiguration steps* (or simply *steps*). In this regard, a multi-step reconfiguration algorithm is proposed in this paper as a compromise between static and dynamic approaches. Unlike the static approaches, the proposed algorithm allows the configuration of the network to change in some steps in order to handle the DG start-up process. However, unlike the dynamic reconfiguration approaches, the number of reconfiguration steps are limited to a predefined value.

This value is determined according to the policies or experience of distribution system operators. More number of stages helps to better procure the restoration objectives. However, as already said in section I, it may cause more switching disturbances to the restored network. The number of steps should be at maximum equal to the number of DGs plus one. Actually, as soon as all the DGs restored are ready to produce no further reconfiguration is allowed.

As illustrated in the previous example, during the DG start-up process, it may happen that the whole unsupplied network is not energized[2] (see Fig. 2.a). This so-called *partial restoration scenario* happens since the DG is not still ready to produce. For such cases, if the loads cannot be detached from their nodes - which is usually the case in practice - there will be no feasible solution for the first configuration step. For these partial restoration scenarios, the authors in [15], [16] proposed to isolate one or some parts of the off-outage area and leave them unrestored. In this respect, the isolated areas are identified using some additional binary variables, which increase the complexity of the problem. This drawback has been already addressed by the authors of this paper proposing a novel model for the reconfiguration problem [17]. This model is based on flow variables and is capable to identify the possible areas to be isolated. However, it could violate the radiality constraint in cases where DGs exist in the off-outage area. This infeasibility is discussed and addressed in section III of this paper.

---

[1] Every node is equipped with a load breaker. Fig. 2 shows only those load breakers that should be operated.

[2] A network is completely energized if all the nodes are under voltage. The loads in an energized network could be either supplied or not.

## III. Problem Formulation

In this section, an analytical formulation is provided for the multi-step reconfiguration problem in the form of a mixed-integer convex optimization problem. The main decision variables of the proposed optimization problem are I) the line switching actions at each reconfiguration step, II) the load switching actions during the whole restorative period, and II) the active/reactive power dispatch of DGs during the whole restorative period. Since the network configuration is fixed between deploying each two successive steps, instead of indexing binary switching variables $Y_{ij}$ with each single time step $t$, they will be indexed with reconfiguration steps $s$. In this way, the number of binary decision variables and therefore the computation burden of the problem decreases drastically.

$$\text{Minimize: } F^{obj} = W_{re}.F^{re} + W_{sw}.F^{sw} + W_{op}.F^{op} \quad (1)$$

$$F^{re} = \sum_{t}\sum_{i\in N} D_i.(1 - L_{i,t}).P_{i,t}^D \quad (2)$$

$$F^{sw} = \sum_{s}\left(\sum_{ij\in W_{tie}^S} Y_{ij,s}.\lambda_{ij} + \sum_{ij\in W_{sec}^S} S_{ij,s}.\lambda_{ij}\right) \quad (3)$$

$$F^{Op} = \sum_{t}\sum_{(i,j)\in W} r_{ij}.F_{ij,t} \quad (4)$$

Subject to:

$$0 \leq Z_{ij,s} \leq 1 \quad \forall ij \in W^S, \forall s \quad (5)$$

$$Z_{ij,s} = Y_{ij,s}, \quad Z_{ji,s} = 0 \quad \forall ij \in W_{ava}^S, \forall s \quad (6)$$

$$Z_{ij,s} + Z_{ji,s} = Y_{ij,s} \quad \forall ij \in W^S, \forall s \quad (7)$$

$$\begin{cases} X_{i,s} = \sum_{j:(i,j)\in W^*} Z_{ji,s} \leq 1, & i \in N^* \\ X_{i,s} = 1, & i \in N\setminus N^* \end{cases} \quad (8)$$

$$X_{ij,s} = \begin{cases} \sum_{k\in Z^*}(A_{ij,k}.X_{k,s}) & : ij \in W^*\setminus W^S \\ Y_{ij,s}, & : ij \in W^S \\ 1, & : ij \in W\setminus W^* \end{cases} \quad \forall s \quad (9)$$

$$0 \leq Y_{ij,s} \leq M.F_{ij,s} \quad \forall ij \in W^S, \forall s \quad (10)$$

$$\begin{aligned} 0 \leq \Psi_{ij,s} \leq M.Z_{ij,s} \\ 0 \leq \Psi_{ji,s} \leq M.Z_{ji,s} \end{aligned} \quad \forall ij \in W^S, \forall s \quad (11)$$

$$\sum_{\forall j^*:(j^*,i)\in W}(\Psi_{j^*i,s}) = \sum_{\forall j^*:(i,j^*)\in W}(\Psi_{ij^*,s}) + X_{i,s} \quad \forall i \in Z^*, \forall s \quad (12)$$

$$\sum_{\forall(i,j)\in W_{ava}^S}\Psi_{ij,s} = \sum_{\forall i\in Z^*} X_{i,s} \quad \forall s \quad (13)$$

$$\begin{cases} (1-Y_{ij,s}) + X_{i,s} - 1 \leq S_{ij,s} \leq 1 \\ (1-Y_{ij,s}) + X_{j,s} - 1 \leq S_{ij,s} \leq 1 \\ 0 \leq S_{ij,s} \end{cases} \quad \forall ij \in W_{sec}^S, \forall s \quad (14)$$

***Time mapping constraints***
***Load Pick-up constraints***
***OPF constraints***

The objective function of the restoration problem ($F^{obj}$) is formulated in (1), which consists of reliability ($F^{re}$), switching ($F^{sw}$) and operational ($F^{op}$) terms, in decreasing order of priority. This hierarchical priority is enabled using $\epsilon-constraint$ method [18]. As formulated in (2), the reliability term expresses the total energy of loads that cannot be restored during the restorative period, while accounting for the load importance factors. The second priority term minimizes the total operation time of line switches including tie switches and sectionalizing switches, which are respectively formulated in the two sub-terms of (3). Since tie switches are normally-open, they are operated when they are energized ($Y_{ij} = 1$). In the case of the sectionalizing switches, their operation has to be determined using auxiliary variables $S_{ij,s}$, and the constraints (14) that are explained later on.

Unlike the first two objective terms, the third term has a very small weight ($W_{op}$) in the objective function. As expressed in (4), the operational term aims to minimize the total active power losses in the restored network. This term is included in the objective function just to ensure the feasibility of the OPF solution binding the squared current variables at the optimal value. It should be noted that the weighting factors are applied on the normalized terms of the expressions defined in (2) and (4).

The network configuration is modeled using constraints (5)-(13), where $M$ is a large multiplier. At each reconfiguration step $s$, the energization status of line $ij$ which is equipped with a switch is identified by a binary variable $Y_{ij,s}$ and its orientation with respect to a virtual source node is determined by continuous variables $Z_{ij,s}$ and $Z_{ji,s}$. This virtual source node is defined as a node in the healthy feeder that is connected to an available tie-switch (ex. node 14 in Fig. 1). If the line is oriented from node $i$ to node $j$, variable $Z_{ij,s}$ will be 1 and $Z_{ji,s}$ will be zero and if the line is oriented from node $j$ to node $i$ variable $Z_{ji,s}$ will be one and $Z_{ij,s}$ will be zero.

For cases with no DG in the off-outage area, an empirical discussion is provided in [17] showing that the constraint set (5)-(10) ensures a radial configuration. However, in cases with DG, this approach could lead to an isolated loop where the loads are supplied in an islanded way using an existing DG in that loop. In order to avoid this, to each line $ij$ with switch in the off-outage area, two continuous flow variables $\Psi_{ij}$ and $\Psi_{ji}$ are assigned. They are associated with the binary variables $Z_{ij}$ and $Z_{ji}$, respectively. The flow $\Psi_{ij}$ or $\Psi_{ji}$ is positive if it is travelling in the same direction as the orientation of the line carrying this flow. As formulated in (11), for each line $ij$ with switch, at most one of the variables $\Psi_{ij,s}$ and $\Psi_{ji,s}$ gets a nonzero value depending on the line orientation that is identified with the variables $Z_{ij,s}$ and $Z_{ji,s}$. Constraint (12) formulates the flow[3] balance equation for each zone[4] $i$, assuming that each zone consumes a flow value equal to one. Finally, (13) implies that the total flows provided by all the available tie-switches as virtual sources must be equal to the total number of energized zones. In Fig. 3, the flow values are indicated for each line with switch. As it can be seen in this example, there is one tie-switch with the flow value of 5 that is energizing 5 zones. It should be mentioned that in order to respect (11)-(13), the lines that are constituting an isolated loop can get only zero flow values. It means that the nodes of this isolated loop will not be energized. Regarding this discussion and the one provided in [17], it can be concluded that (5)-(10) ensure a radial configuration without creating isolated loops supplied by DGs in islanded mode.

---

[3] It is meant directional flow with regard to the graph theory and should be distinguished with the electrical flow term used in the OPF formulation

[4] In this paper, a zone is referring to each segment of the feeder that is surrounded by two or more sectionalizing switches (Fig. 1).

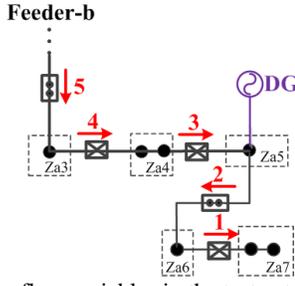

Fig. 3. The auxiliary flow variables in the test network of Fig. 2.b.

The operation of sectionalizing switches is formulated in (14). Unlike tie-switches, for sectionalizing switches, energization status does not necessarily imply that they should be operated or not. At a given step $s$, the sectionalizing switch on line $ij$ must be operated (opened) ($S_{ij,s} = 1$) only if it is de-energized ($Y_{ij,s} = 0$) and at least one of its ending buses are energized ($X_{i,s} = 1$ or $X_{j,s} = 1$). According to (14), for the other sectionalizing switches, $S_{ij,s}$ is free to get any value between 0 and 1. However, since the sum of $S_{ij}$ variables is minimized according to (3), $S_{ij}$ for these switches will be zero, meaning that they should not be opened.

### A. Time mapping Constraints

As mentioned earlier, in order to reduce the number of binary decision variables, reconfiguration variables are indexed with each reconfiguration step. Now in order to ensure that the operational safety constraints are met, the temporal dependency of the reconfiguration variables should be correlated with the ones of the OPF variables. For this aim and since OPF variables are indexed with each time slot, each step is mapped to the time slots, during which its corresponding configuration is under operation. This is determined using variable $K_{t,s}$ under constraints formulated in (15). Being of SOS1-type, constraint (15.a) ensures that for each time slot $t$, variable $K_{t,s}$ is 1 only for one reconfiguration step $s$ and is 0 for the rest. In other words, time slot $t$ can be assigned only to one step according to the following condition:

$$\text{if } T_s \leq t < T_{s+1} \text{ holds, then } K_{t,s} = 1 \text{ must hold}$$

This conditional constraint is equivalent to the following expression:

$$(T_s > t) \text{ or } (t \geq T_{s+1}) \text{ or } (K_{t,s} = 1) \text{ must hold}$$

It means among the three constraints given above, only one must hold. Since $K_{t,s}$ is a binary variable, one can conclude that the first two constraints could hold only when $K_{t,s} = 0$ holds. This expression is formulated in (15.b-c). According to (15.b), $T_s > t$ could hold only when $K_{t,s} = 0$, and according to (15.c) $t \geq T_{s+1}$ could hold only when $K_{t,s} = 0$.

$$\begin{cases} \sum_{s^* \in N_s} K_{t,s^*} = 1, \quad :SOS1 & (15.a) \\ T_s - M(1 - K_{t,s}) \leq t & \forall t,s \quad (15.b) \\ t \leq T_{s+1} + M(1 - K_{t,s}) & (15.c) \end{cases}$$

$$\sum_{ij \in W_{tie}^S} Y_{ij,s}.\lambda_{ij} + \sum_{ij \in W_{sec}^S} S_{ij,s}.\lambda_{ij} + T_{s-1} \leq T_s \quad \forall s \quad (16)$$

Constraint (16) determines the time instant of each reconfiguration step. The network undergoes the configuration at step $s - 1$ starting from $T_{s-1}$ until $T_s$, assuming that $T_0 = 0$. This duration must be larger than the time needed to deploy the switching operation at step $s$. Since at a given step s, the switching actions are deployed in a successive way, their deployment times are summed up in (16) resulting in the switching operation time of step s. $\lambda_{ij}$ is pre-determined according to the average time that is taken to operate switch $ij$. This average time for the case of manually-controlled switches depends among others on the distance from the switch location to the operation center.

### B. Load Pick-Up Constraints

Regarding the partial restoration scenarios, along with the possibility to setup isolated areas, the loads that are equipped with a breaker at their nodes could be shed to respect the safety constraints particularly at the early stages of the network reconfiguration. If feasible, it is scheduled to pick-up these loads during the rest of the restorative period. In this regard, at each time slot t belonging to reconfiguration step s ($K_{t,s} = 1$), a decision is made with binary variable $L_{i,t}$ in (17) for a given energized node ($X_{i,s} = 1$), indicating if its load will be supplied or rejected (1/0). It is assumed that once a given load in the off-outage area is restored at a given time, no further interruption is permitted during the subsequent time slots of the restorative period. This constraint is formulated in (18).

$$L_{i,t} \leq X_{i,s} + 1 - K_{s,t} \qquad \forall i \in N^*, \forall t,s \quad (17)$$

$$\begin{cases} 0 \leq L_{i,t-1} \leq L_{i,t} \leq 1, & \forall i \in N^*, \forall t \\ L_{i,t} = 1, & \forall i \in N \setminus N^*, \forall t \end{cases} \quad (18)$$

### C. OPF Constraints

In this sub section, a relaxed AC-power flow formulation that is proposed in [17] is integrated to the optimization problem. The aim is to ensure the feasibility of the solution concerning the operational safety constraints.

$$0 \leq F_{ij,t} \leq f_{ij}^{max\,2}(X_{ij,s} + 1 - K_{s,t}) \qquad \forall ij \in W, \forall t,s \quad (19)$$

$$\begin{aligned} -M(X_{ij,s} + 1 - K_{s,t}) &\leq p_{ij,t} \leq M.(X_{ij,s} + 1 - K_{s,t}) \\ -M(X_{ij,s} + 1 - K_{s,t}) &\leq q_{ij,t} \leq M.(X_{ij,s} + 1 - K_{s,t}) \\ &\forall ij \in W, \forall t,s \quad (20) \end{aligned}$$

$$v^{min\,2}(X_{i,s} - 1 + K_{s,t}) \leq V_{i,t} \leq v^{max\,2}(X_{i,s} + 1 - K_{s,t})$$
$$\forall i \in N, \forall t,s \quad (21)$$

$$-M.(2 - X_{ij,s} - K_{s,t}) \leq V_{i,t} - V_{j,t} - 2(r_{ij}.p_{ij,t} + x_{ij}.q_{ij,t})$$
$$\leq M.(2 - X_{ij,s} - K_{s,t}) \quad \forall (i,j) \in W, \forall t,s \quad (22)$$

$$p_{ij,t} = (\sum_{\substack{i^* \neq i \\ i^*j \in W}} p_{ji^*,t}) + r_{ij}.F_{ij,t} + L_{i,t}.P_{j,t}^D - P_{j,t}^{inj}$$
$$\qquad\qquad\qquad\qquad\qquad\qquad\qquad (23)$$
$$\forall ij \in W, \forall t,s$$

$$q_{ij,t} = (\sum_{\substack{i^* \neq i \\ i^*j \in W}} q_{ji^*,t}) + x_{ij}.F_{ij,t} + L_{i,t}.Q_{j,t}^D - Q_{j,t}^{inj}$$

$$\forall ij \in W, \forall t, s \quad (24)$$

$$\left\| \begin{matrix} 2p_{ij,t} \\ 2q_{ij,t} \\ F_{ij,t} - V_{i,t} \end{matrix} \right\|_2 \leq F_{ij,t} + V_{i,t} \quad \forall ij \in W, \forall t, s \quad (25)$$

$$\left\| \begin{matrix} P_{i,t}^{inj} \\ Q_{i,t}^{inj} \end{matrix} \right\|_2 \leq S_{i,max}^{inj} \quad \begin{matrix} \forall i \in \Omega_{Res} \\ \forall t, s \end{matrix} \quad (26)$$

$$E_{i,t} = E_{i,t-1} - P_{i,t}^{inj} \Delta t \geq 0 \quad \forall i \in \Omega_{DG}, \forall t, s \quad (27)$$

$$0 \leq P_{i,t}^{inj} \leq M(X_{i,s} + 1 - K_{s,t})$$
$$-(X_{i,s} + 1 - K_{s,t}) \leq Q_{i,t}^{inj} \leq M(X_{i,s} + 1 - K_{s,t})$$
$$\forall i \in \Omega_{DG}, \forall t, s \quad (28)$$

$$0 \leq P_{i,t}^{inj} \leq P_{i,max}^{inj}(2 - X_{i,s} - K_{s,t} + K_{s,t-\Delta_i})$$
$$Q_{i,min}^{inj}(2 - X_{i,s} - K_{s,t} + K_{s,t-\Delta_i}) \leq Q_{i,t}^{inj}$$
$$\leq Q_{i,max}^{inj}(2 - X_{i,s} - K_{s,t} + K_{s,t-\Delta_i})$$
$$\forall i \in \Omega_{DG}, \forall t > \Delta_i, \forall s \quad (29)$$

In (19)-(22), the discrete variables of the multi-step reconfiguration problem ($X_{i,s}, X_{ij,s}$) are linked to the operational variables of the OPF problem using variable $K_{t,s}$. Therefore, the operational variables at time $t$ follow the network configuration step $s$ that is under operation at that time ($K_{t,s} = 1$). It is worth to note that the configurations at other steps $s^*$ ($K_{t,s^*} = 0$) have no impact on the OPF variables at time $t$.

Constraints (19) and (20) enforce, respectively, current and active/reactive power flows to be zero for each de-energized line ($X_{ij,s} = 0$). The squared voltage magnitude is set to zero in (21), for de-energized nodes ($X_{i,s} = 0$), and kept within the feasible region for energized nodes ($X_{i,s} = 1$). Constraint (22) expresses the nodal voltage equations only for the energized lines ($X_{ij,s} = 1$). The active and reactive power balance equations at the ending node of each line are formulated in (23) and (24), respectively. Constraint (25) is the relaxed version of the current flow equation in each line as proposed in [17].

The apparent power injection of DGs and substations is limited by the cone constraint of (26). In this paper, it is assumed that the reservoir (e.g. gas tank or diesel tank) supplying the DG primary resource has a finite energy capacity during the restorative period. The availability of energy stored at the reservoir of each DG at time $t$ is ensured according to (27). The active-/reactive power injection of DGs at time $t$ are forced in (28) to be zero if the hosting node is not energized according to the network configuration at time $t$. As mentioned in section I and formulated in (29), for a given DG in the off-outage area, once its hosting node is energized ($K_{s,t} = 1, X_{i,s} = 1$), it still cannot inject power before it is fully restarted ($K_{s,t-\Delta_i} = 0$). This restart process is assumed to take $\Delta_i$ time steps.

## IV. SIMULATION AND DISCUSSION

The proposed analytical model for the restoration problem is applied on a 11.4 kV balanced test network based on a practical distribution network in Taiwan. As shown in Fig. 4, this test system includes 2 substations, 11 feeders, 84 nodes and 94 branches (incl. tie-branches). The details regarding the nodal and line data are given in [19]. The base power and energy values are set to 1MW and 1 MWh, respectively. This network hosts 4 dispatchable DGs. The active/apparent power capacities of DGs installed on nodes 7, 39, and 80 are equal to 2.8MW/3MVA, whereas the capacities of the DG at node 59 are equal to 0.8MW/1.0MVA. The start-up duration of these DGs is assumed to be 30 minutes for each. Besides these dispatchable DGs, there exists also non-dispatchable DGs including a PV and a Wind generator at nodes 28 and 45, respectively. These intermittent DGs are modeled as voltage-independent power injection units with zero reactive power components. In this paper, their respective forecasted active power generation is derived from the real data reported in [20] and [21]. The minimum and maximum voltage magnitude limits are set, respectively, to 0.917 and 1.050 p.u. [22].

The active/reactive profiles of rural, light, and high industrial loads reported in [23] are assigned to the loads in the network of Fig. 4 according to [17]. It is assumed that each node of the network shown in Fig. 4 is equipped with a load breaker. The manually-controlled line switches are labeled with '*', while the rest together with all the load breakers are remotely-controlled. It is assumed that the time needed for the operation of each manually controlled and remotely-controlled switch is 30 and 0.5 minutes, respectively. The critical loads, that are shown with '⋈' in Fig. 4, have priority factors equal to 100 while the priority factors of the other loads are equal to 1. The algorithm is implemented on a PC with an Intel(R) Xeon(R) CPU and 6 GB RAM; and solved in Matlab/Yalmip environment, using Gurobi solver. Branch-and-Bound method is used to handle the developed mixed-integer optimization problem. The restorative period is assumed from 9:00 to 15:00. The time resolution of this study is chosen to be 15 minutes.

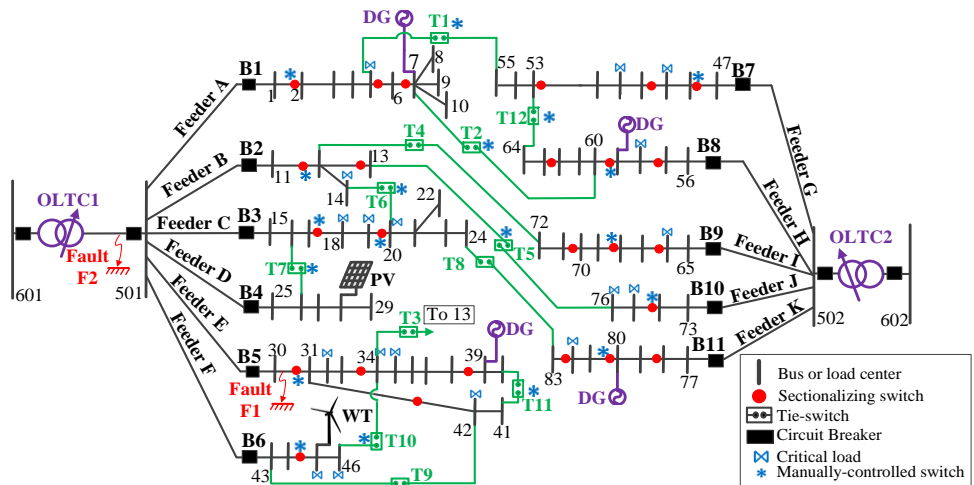

Fig. 4. Test distribution network in normal state configuration [19].

Table I. Optimal restoration results in case of fault scenarios 1 and 2.

| Scenarios | Step Number | Switching actions | $F^{sw}$ (min) | $F^{re}$ (p.u.) | Computation time (min) | Min. voltage (p.u.) | Max. voltage (p.u.) | Min. current margin (p.u.) |
|---|---|---|---|---|---|---|---|---|
| 1.a | Step1 | I. Open load breakers {31,34,41} and switch 33-34 II. Close {T3,T9} | 3 | 8.3 | 1.16 | 0.9570 p.u. at node 38 | 1.05 p.u. at node 39 | 0.5192 p.u. on line 501-43 |
| | Step2 | II. Close T11 III. Open switch 38-39 | 30.5 | | | | | |
| 1.b | Step1 | I. Open switch 31-42 and load breakers {31,34,37} II. Close {T3,T9} | 3 | 14.63 | 0.68 | 0.9375 p.u. at node 38 | 1.05 p.u. at node 501 | 0.0062 p.u. on line 11-12 |
| 2 | Step1 | I. Open feeder breakers{B3,B4,B5,B6}, switch 12-13, and load breakers{3,4,6,7,9,10,12,20} II. Close {T4,T8} | 7.5 | 664.98 | 10.46 | 0.9219 p.u. at node 46 | 1.05 p.u. at node 59 | 2.82 p.u. on line 502-65 |
| | Step2 | I. Open switch 31-42 and load breakers {13,31,33,34,35,36,37,38,41,42,44} II. Close {T3,T9}, feeder breaker B6,and switch12-13 | 8 | | | | | |

*A. Scenario1: comparison of multi-step reconfiguration with static and dynamic approaches*

In this scenario, the restoration strategy is found in case of fault F1 shown in Fig. 4. In this regard, first, the developed multi-step restoration strategy is tested in scenario 1a, where the complete service restoration is deployed through two reconfiguration steps. The switching operations at each step are reported in Table I according to the order they should be deployed. The sketch of the timing of the restoration plan is shown in Fig. 5.a. Following Fault F1, the dispatchable DG at node 39 is disconnected. In order to energize this DG as soon as possible, its hosting node is energized after the first step, relying only on the remotely controlled switches. As it can be seen in Fig. 5.a, the loads at buses 31 and 34 can be restored only after t=09:30 A.M., when the DG starts to inject power.

In the next step, the dynamic reconfiguration method is applied to the simulation scenario 1.a. In this regard, it is assumed that the configuration of the network can be changed after each 30 minutes (two time steps). Therefore, binary switching variables $Y_{ij}$ are indexed with each of these considered time steps. The obtained restoration solution is the same as the one given in Table I. This is due to the large weight of the reliability term in the objective function. However, the computation time with dynamic approach is equal to 24.11 minutes which is significantly longer than the one obtained with the multi-step reconfiguration approach. This shows that using the proposed multi-step reconfiguration method, the computational burden of the dynamic formulation is relieved without compromising the quality of the solution.

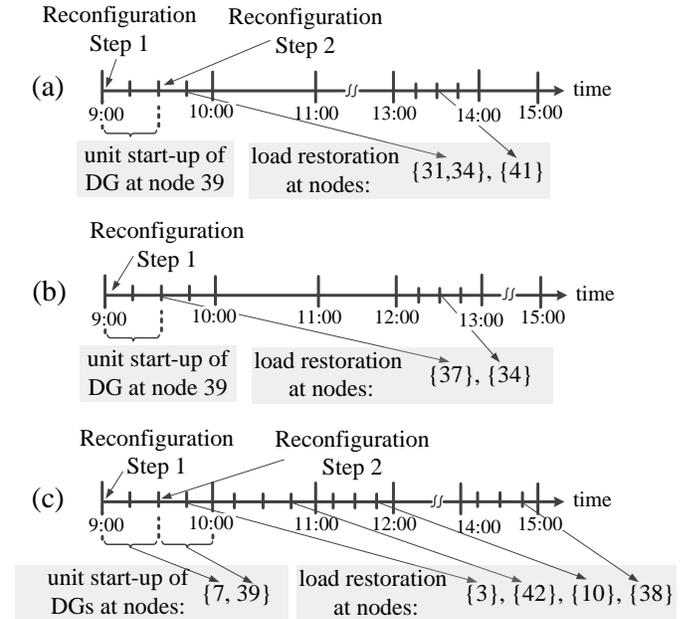

Fig. 5. The sketch of the timing of the restoration plan in a) scenario 1.a, b) scenario 1.b., and c) scenario 2.

In scenario 1.b, the test case of the scenario 1.a is studied again having only a single reconfiguration step. As the results of Table I shows, compared with the two-step reconfiguration, having a single reconfiguration step degrades significantly the quality of the restoration solution in terms of the reliability objective value ($F^{re}$). The sequence of DG starting and load pickup is shown in Fig. 5.b. As it can be seen, the load at node 31 that was detached at the beginning of the restoration strategy (see Table I) cannot be picked up having only one reconfiguration step.

*B. Scenario2: a large off-outage area*

In this scenario, a fault occurs on substation 601 (fault F2 in Fig. 4), making all the feeders connected to substation 601 unsupplied. The scalability of the proposed multi-step formulation in comparison with the dynamic formulation is better understood in this scenario with such a huge outage area. The dynamic formulation in this scenario ends up with 1521 binary variables. The available computing facilities cannot solve this complex optimization problem and it failed to provide even a single feasible solution. However, with the proposed multi-step formulation, the number of binary variables is reduced to 396 variables. As reported in Table I, the optimal solution is obtained with the same computing facilities in 10.46 minutes only. It shows that the proposed formulation can be applied efficiently to large scale outages in distribution networks.

As reported in Table I, the reliability objective value is equal to 664.98 p.u. . This value increases to 967 p.u. if only a single reconfiguration step is allowed. This shows that the proposed multi-step reconfiguration problem improves the quality of the restoration problem with respect to the static approaches especially in cases with large off-outage area. The switching operations of each reconfiguration step are reported in Table I and the timing sketch of this restoration plan is shown in Fig. 5.c. As mentioned in section I, the timing of reconfiguration steps is optimized according to the optimal sequence of DG starting and load variation. The switching operations and the resulting configuration of each step are depicted in Fig. 6. Under the configuration at step 1, the parts in the off-outage area that are coloured with purple and blue are restored through available tie-switches T4 and T8, respectively. Nodes 7 and 39 that are hosting unsupplied DG are energized at steps 1 and 2, respectively. However due to the DG energization time constants, both of these DGs start to inject power only when the network is operated under the configuration at step 2. Therefore, during this time period, some of the unrestored loads can be picked up as illustrated in Fig. 5.c. However, feeder D is still left isolated without any supply until the end of the restorative period. The possibility to have such a partial restoration solution is missing in the literature, while respecting the radiality in presence of DGs (section I).

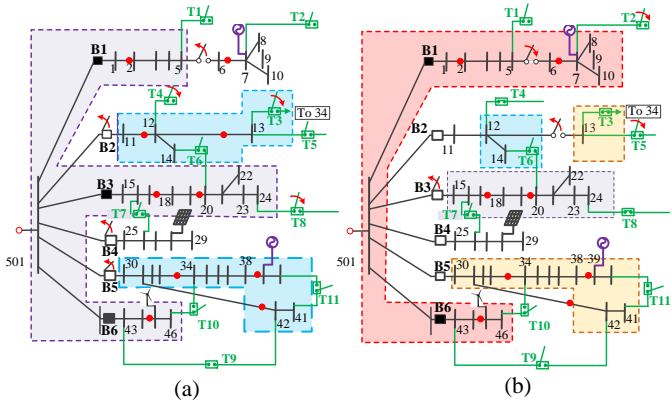

Fig. 6. The optimal reconfiguration scheme of simulation scenario 2 at a) step 1 and b) step2.

In order to validate the feasibility of the obtained solution regarding the safety constraints, the real values of voltage and current profiles are derived using post power flow simulations running for the obtained new configurations of the network during the whole restorative period. These power flow simulations are made using Matlab/MATPOWER toolbox. The representative numerical results are given in Table I for each simulation scenario.

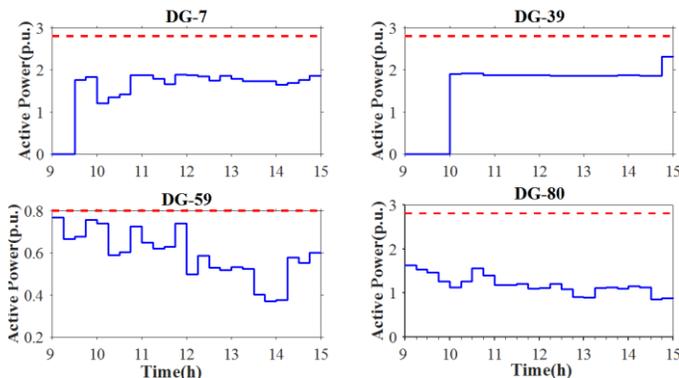

Fig. 7. The active power dispatch of DGs and their limits in simulation scenario 2.

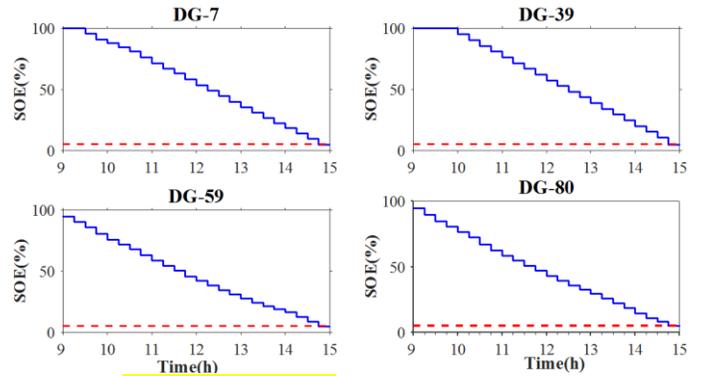

Fig. 8. The state of energy (SOE) of DGs and their limits in simulation scenario 2.

Fig. 7 shows the optimal DG set points according to the restoration solution. It illustrated in Fig. 7 and Fig. 8, how the DGs contribute to the proposed restoration strategy within their power and energy capacity envelopes, respectively. As it can be seen, although nodes 7 and 39 are energized, respectively, at times 9:00 and 9:30, their hosted DGs start to inject power, respectively, from times 9:30 and 10:00 on. These results show that the network and DGs are operated within the safe region but very close to the limits in order to benefit at best of their capacities for the load restoration.

## V. CONCLUSION

In this paper, a multi-step model is proposed for the reconfiguration problem as a compromise between the static and dynamic approaches, which already exist in the literature. This model is applied to the restoration problem resulting in a MISOCP optimization problem. The start-up requirements of grid-connected DGs, including the energization of their hosting nodes and their start-up duration s are accurately modeled and integrated to the proposed formulation for the restoration problem. In this regard, a new formulation is proposed for modeling the radiality constraints in presence of DGs which also account for partial restoration scenarios. In order to illustrate different features of this proposed contribution, the developed model is successfully tested on a 70-bus distribution network through two simulation scenarios. As shown in the simulation and discussion section, the timing schedule of the different phases of the restoration plan is optimally determined within a reasonable computation time even for a large-scale test case. The superiority of the proposed multi-step formulation with respect to the existing static and dynamic formulation is validated using these two simulation scenarios.

The application of the proposed multi-step reconfiguration problem is not limited to consider only the starting process of DGs in the service restoration. This formulation is applicable when it is desired to consider the effect of any time-dependent process on the solution of the reconfiguration problem.

## VI. ACKNOWLEDGMENT

The authors gratefully acknowledge the financial support of the Qatar Environment and Energy Research Institute (QEERI).